# Comment on "Dimensionless Units in the SI"

Paul Quincey

Analytical Science Division, National Physical Laboratory, Hampton Road, Teddington, TW11 0LW, United Kingdom.

Tel: +44 (0)20 8943 6788; E-mail address paul.quincey@npl.co.uk

**Abstract**

The recent paper by Mohr and Phillips (2015) describes several problems relating to the treatment of angle measurement within SI, the unit hertz, and quantities that can be considered countable (rather than measureable). However, the proposals that they put forward bring new problems of their own. This paper proposes alternative suggestions that solve the problems less painfully. Specifically, clarifying the text on angle in the SI brochure; relegating the hertz to a "Non-SI unit accepted for use with the International System of Units", with specific application only for "revolutions or cycles per second"; and encouraging countable quantities to be presented as pure numbers, while requiring that a sufficient description of the quantity being counted is given in the accompanying text.

**Introduction**

The current revision of several base units[1] within SI (Milton et al 2014) has encouraged a more general review of SI units as part of the revision of the SI Brochure (BIPM 2006). One of the key areas in this review is the treatment within SI of dimensionless quantities, including plane and solid angle, and countable quantities like atoms or cycles of periodic phenomena. There are specific deficiencies that need to be addressed, such as the fact that the brochure does not actually define the size of a radian or steradian (beyond the unhelpful statement "the radian and steradian are special names for the number one"), and the more important problem that the unit hertz is defined as $s^{-1}$, and is widely used to mean one cycle per second, when the coherent SI unit $s^{-1}$ as applied to cyclical frequency is one radian per second.

---

[1] While the revision of the SI in terms of defining constants can be seen as removing the need to distinguish between base units and derived units, the distinction is nevertheless a useful one, and it is used extensively in the draft 9th edition of the SI Brochure, which describes the revised SI, available at file://fpsvr2/users5$/pq/Downloads/si_brochure_draft_ch123%20(5).pdf

There has been little published discussion of most of these issues since the 1980s. The recent paper by Mohr and Phillips (2015) – hereinafter referred to as MP - proposes some rather radical changes, including the treatment of radians and steradians as something very similar to new base units, which would indirectly help with the hertz problem, and the introduction of extra unit symbols for countable quantities. The aim of this paper is to point out some of the problems with these proposals, and to put some alternative proposals into the discussion.

**The problem with treating the radian as a base unit**

The status of the quantities plane angle and solid angle within any system of measurement has genuine subtleties to it, as shown by their original, unique, position in the SI of 1960 as supplementary units, before they were declared derived SI units in 1995. The slow evolution of the SI approach to this arcane field means that it is easy for the underlying arguments to be forgotten.

At the heart of the problem is the fact that there are two natural units for angle. For mathematicians and theoretical physicists it is the radian, while for everyone else it is the revolution. To say "rotate this lever by 1 natural unit of angle" is clearly ambiguous without an agreed convention. If the unit is not clear, an erroneous factor of 2π will appear when the "wrong" unit is chosen. The mathematical convenience of the radian means that there is no real question about whether the radian or the revolution (or some subdivision of it) should be the chosen unit for angle within the SI – it should be the radian - although this will not be evident to users of protractors or theodolites.

One way to avoid the ambiguity is to give the unit for angle the status of a base unit – making the unit radian explicit at all times. MP does not actually use the words "base unit", but their recommendation that the radian "should be regarded as the coherent unit for angles in the SI" and relevant quantities should be "reported including radians as a unit", with consequent changes to the units to be used for various fundamental constants, seems to amount to much the same thing. It is not clear from the paper whether it is proposed that the radian should be treated as a base unit, a derived unit, and if so in what way it is derived from the base units, or something else.

While there is some freedom of choice in the number of base units that can be used, shown for example by the addition of the mole as a base unit in 1971, the choice of radian as a base unit brings unexpected consequences. One of these is that there needs to be a distinction between "length-type" lengths, measured in metres, and "radius-type" lengths, measured in metres per radian. The paper by Eder (1982), with which MP implicitly agrees, showed that it is possible for this system to be used consistently, provided some changes are made to the normal way of doing things. Some of these changes are described in MP, affecting, for example, the units for many fundamental constants. MP suggests that most situations would be covered by applying a "rule of thumb" that replaces instances of "radius" (with the unit m) in expressions with "angular radius of curvature" (with the unit m rad$^{-1}$).

One example given is the Bohr radius, which would have the unit m rad$^{-1}$. This immediately raises the question of which units should be used to describe the size of an atom. Would we say that the size of a gaseous gold atom is measured in m rad$^{-1}$, while the spacing of gold atoms in a solid lattice is measured in m (the magnitude of the values being curiously similar)? And what about the correct unit for the size of a gold atom in liquid gold – at what temperature does the phase change for the unit take place? This brings significant complications to what should be a very simple concept. If students are to be taught that the Bohr radius is a measure of the angular radius of curvature of an atom, not its size, there would need to be a very good reason for this.

There are worse problems when it comes to dimensional analysis. Applying the MP rule of thumb, you would think that the units for the expression $\pi r^2$ are m$^2$ rad$^{-2}$. It is explained in Eder that the formula for the area of a circle, $\pi r^2$, gives an answer in m$^2$ because we are to consider one r to have the unit m rad$^{-1}$, the other r to have the unit m, and $\pi$ to have the unit rad, confirming that the rule of thumb does not work even in this simple situation. And we cannot just invent an extra rule of thumb that "r$^2$" always has units m$^2$ rad$^{-1}$, because in the formula for moment of inertia, "r$^2$" has units m$^2$ rad$^{-2}$.

This effectively invalidates the valuable technique of dimensional analysis as a way of checking that the product of an expression has the right physical dimensions. Take as an example the expression for Stokes drag force, $3\pi\eta vd$, where $\eta$ is viscosity (unit N.s.m$^{-2}$), v is velocity, and d is diameter. In the current set of base units it is immediately apparent that this expression gives an answer in the expected units of newtons. If the radian is considered to be a base unit, we would need to decide whether the diameter should be considered a "length-type" or a "radius-type" quantity, and also whether the π was representing an angle, and therefore considered to have the unit radians – neither of which is evident from the expression itself.

**Plane angle and solid angle – a way forward**

There are, then, significant practical problems with treating the radian and steradian as base units, even though this would, in many cases, remove the potential ambiguity between radians and revolutions. The present simplistic two-category system means that the alternative is for them to be derived units, but derived units that are dimensionless. I will try to show that there are problems with treating angles in the same way as other dimensionless quantities, but that these can be effectively overcome simply by providing a clearer explanation.

MP states (in Section 3) that plane angle can be shown by mathematical reasoning to be either dimensionless or having the dimension of length squared. It says that "this illustrates that conclusions about the dimensions of quantities based on such reasoning are clearly nonsense." However, the example given is flawed. The definition of an angle in radians as "twice the area of the sector which the angle cuts off from a unit circle whose centre is at the vertex of the angle" is entirely correct. However, the term "unit circle" is from the world of mathematics rather than metrology - in the real world, a circle cannot have a radius of exactly one metre by declaration. Metrologically speaking, the angle is here defined as twice the measured area of the sector, which in SI is expressed as square metres, divided by the "unit area", which is the measured radius of the circle, in metres, squared. Put algebraically, this approach defines the radian by using the expression for the area of a sector, $A = ½r^2θ$, where A is the area, r is the radius, and θ is the angle in radians. Angle is therefore dimensionless in this case, as in the usual "arc-length divided by radius" case. Conclusions based on such reasoning may not be nonsense after all.

What this point does show is that plane angle can be seen as having dimensions of $m^2\,m^{-2}$, as well as $m\,m^{-1}$. It is dimensionless, but it is not correct to identify a radian with a $m\,m^{-1}$ (or a steradian with a $m^2\,m^{-2}$).

The current line in the SI brochure is, more or less, that all dimensionless quantities in effect have the unit "one", and that is all you need to know about them to interpret results. Countable quantities, together with simple ratios like refractive index, are dimensionless, and for them this line is fair enough. However, plane angle and solid angle, although best treated as dimensionless, are different. Statements in the brochure like "the radian and steradian are special names for the number one", are both conceptually dubious and of no help to users. The degree is a commonly-used unit for plane angle, accepted for use with the SI, but it would be silly and unhelpful to state that "the degree is a special name for the number π/180."

Neither "the number one" nor "a metre per metre" gives a clear definition of a radian. As an example of the latter point, we could choose to define angle as the ratio of the length of arc of a circle to its circumference (rather than its radius). This would be a valid measure of angle, in units "metre per metre", but one that gives the answer in revolutions instead of radians.

The actual position is that plane angle and solid angle cannot realistically be treated as base units, but neither are they derived units in the strict sense, because the quantities m/m or $m^2/m^2$, while having some indicative value, do not in themselves define the units radian and steradian. Nor should they be treated the same as "countable" quantities simply because they are dimensionless, because there is more than one natural unit. How this is dealt with in the SI brochure is largely a matter of taste. Rather than reinvent a separate category such as "supplementary units", it is proposed here that the table of "Coherent derived units" is renamed "Angles and coherent derived units". In this table angles would still be treated as dimensionless, but without the "definitions" m/m or $m^2/m^2$. They just need more careful description than is given at present, and this can be achieved by inserting a few lines, such as:

*"The radian is the plane angle subtended at the centre of a circle by an arc equal in length to its radius. The steradian is the solid angle subtended at the centre of a sphere by an area on its surface equal to its radius squared. As a consequence of their special status as the natural units for angle within mathematics, these dimensionless units can formally be replaced within the SI by the number one. In practice the symbols rad and sr are used where appropriate, especially where there is any risk of confusion that units such as degrees or revolutions are being used."*

**What to do with the hertz?**

The problem with the unit hertz is that it often refers to cycles (or revolutions) per second, when the coherent SI unit for cyclical frequency is radians per second. The measure $s^{-1}$ as applied to a cyclical quantity should properly be interpreted within SI as radians per second. It is not acceptable that the unit Hz can be interpreted as radians per second, as is currently the case, when it is intended to represent cycles per second.

Although the hertz problem features strongly in MP (notably in the opening parable), it is not clear what their recommendation in this area is. The paper concludes that "the unit hertz cannot be regarded as a coherent unit of the SI," but it does not say whether this should lead to its use being forbidden within the SI - perhaps to be replaced with unambiguous units such as cycles per second - or some other course of action.

Given that the unit hertz is widely used for wave frequency and therefore here to stay, the logical solution would be simply to remove hertz from the list of coherent derived units and make it, like the degree, a "Non-SI unit accepted for use with the International System of Units", with specific application only for "revolutions or cycles per second", and a note saying that the change in its status has been made to avoid confusion with the coherent derived unit for cyclical frequency, which is rad $s^{-1}$. The key element is to remove the identification of the unit hertz with $s^{-1}$.

**Symbols for countable quantities**

Countable quantities occur in many different situations, such as in describing the concentration of airborne particles in a clean room, which can be expressed in words as the number of particles per cubic metre. The problem arises when trying to express such countable quantities in abbreviated units, because the counted entity can be many distinct things. The MP paper suggests several of these units: evt, cnt, dcy, ent, mcl, atm, and pcl. The problem with this approach soon becomes apparent; if every countable quantity is given a symbol, the SI system of units will become very messy, continually adding new symbols that require international agreement. If the symbol is supposed to give a complete description of what has been counted, we would soon need to employ new alphabets, and the approach would become unworkable.

Underlying the discussions in this area there seems to be a kind of numerophobia – a fear of pure numbers being used as results within the SI. This is not helped by the opening line of the SI brochure: "The value of a quantity is generally expressed as the product of a number and a unit." This is entirely true for dimensioned quantities, but if it leads to problems when forced to apply to dimensionless quantities, it is the mantra that needs to be modified, not good sense.

The question seems to be: if there are no explicit units, how can a result be said to be expressed in SI units? The answer is that quantities validly expressed as pure numbers are automatically not only in SI units, but also in every other system of units, at the same time. They do not depend on the definitions of the SI units, and are independent of whatever system of units is being used. This should be seen as entirely helpful - it is not some sort of problem that the SI needs to address.

It is proposed that, instead of creating a long list of unit symbols for countable quantities, or forcing in a "1" for semantic reasons, with countable quantities the number is left to stand by itself. Whenever countable quantities are expressed within results, a fit-for-purpose set of words describing the quantity must be included. For example, instead of stating that the concentration of oxygen in a room is $10^{25}$ atm m$^{-3}$, and relying on the unit to explain what is meant, it should be stated that the concentration of oxygen atoms in a room is $10^{25}$ m$^{-3}$. This would encourage clarity in the text while removing the temptation to create a new "unit" for every occasion.

To end with another, simpler example: how many planets, excluding dwarf planets, are there in the solar system? The answer is: 8. But what is that in SI units? Is it 8 1; 8 uno; 8 ped; or 8 **PNT** ? The answer should still simply be: 8.

**Acknowledgements**

I would like to thank Peter Mohr, Richard Brown and Stephen Lea for stimulating discussions on these topics.